\newcommand{\version}{January/30/2013}
\documentclass[pra,aps,twocolumn,showpacs]{revtex4}

\usepackage[english]{babel}
\usepackage{latexsym}

\usepackage{amsmath}

\usepackage{amsfonts}

\usepackage{amssymb}
\usepackage{amscd}
\usepackage{amsgen,amstext,amsbsy,amsopn}
\usepackage{math rsfs}

\usepackage{hyperref}

\newcommand{\intN}{\I_N}

\newcommand{\PKerp}{P_{\mathrm{Ker} (\intN) ^{\perp}}}

\newcommand{\muNm}{\mu_{N}}
\newcommand{\muNmone}{\mu_{N} ^{(1)}}
\newcommand{\ZNm}{\mathcal Z_{N}}

\newcommand{\HNm}{H_{N}}

\newcommand{\MFmf}{\E ^{\rm MF}}

\newcommand{\rhoMFm}{\varrho ^{\rm MF}}

\newcommand{\MFel}{\E ^{\rm el}}
\newcommand{\MFth}{\E ^{\rm th}}
\newcommand{\rhoMFel}{\varrho ^{\rm el}}
\newcommand{\rhoMFth}{\varrho ^{\rm th}}

\newcommand{\I}{\mathcal{I}}

\newcommand{\om}{\omega}

\newcommand{\intR}{\int_{\R ^2}}

\newcommand{\bdm}{\begin{displaymath}}
\newcommand{\edm}{\end{displaymath}}
\newcommand{\bdn}{\begin{eqnarray}}
\newcommand{\edn}{\end{eqnarray}}
\newcommand{\bay}{\begin{array}{c}}
\newcommand{\eay}{\end{array}}
\newcommand{\ben}{\begin{enumerate}}
\newcommand{\een}{\end{enumerate}}
\newcommand{\beq }{\begin{equation}}
\newcommand{\eeq }{\end{equation}}

\newcommand{\R}{\mathbb{R}}

\newcommand{\half}{\hbox{$\frac12$}}

\newcommand{\mopt}{m_{\rm opt}}

\newtheorem{teo}{Theorem}

\newcommand{\E}{\mathcal{E}}

\numberwithin{equation}{section}

\begin{document}

\title{Quantum Hall states of bosons in rotating anharmonic traps}

\author{Nicolas Rougerie}
\affiliation{Universit\'e Grenoble 1 \& CNRS,  LPMMC (UMR 5493), B.P. 166, 38 042 Grenoble,
France.}
\author{Sylvia Serfaty}\affiliation{UPMC Univ Paris 06, UMR 7598 Laboratoire Jacques-Louis Lions,
Paris, F-75005 France ;
CNRS, UMR 7598 LJLL, Paris, F-75005 France}
\affiliation{Courant Institute, New York University, 251 Mercer st, NY NY 10012, USA}
\author{Jakob Yngvason}
\affiliation{Faculty of Physics,
University of Vienna, Boltzmanngasse 5
}
\affiliation{Erwin Schr{\"o}dinger Institute for Mathematical Physics, Boltzmanngasse 9, A-1090 Vienna, Austria.}

\date{\version}

\begin{abstract} 
We study a model of bosons in the lowest Landau level in a rotating trap where the confinement potential is a sum of a quadratic and a quartic term. The quartic term improves the stability of the system against centrifugal deconfinement and allows to consider rotation frequencies beyond the frequency of the quadratic part. The interactions between particles are modeled by a Dirac delta potential. We derive rigorously conditions for  ground states of the system to be strongly correlated in the sense that they are confined to the kernel of the interaction operator, and thus contain the correlations of the Bose-Laughlin state. Rigorous angular momentum estimates and trial state arguments indicate a transition from a pure Laughlin state to a state containing in addition a giant vortex at the center of the trap (Laughlin quasi-hole). There are also indications of a second transition where the density changes from a flat profile in a disc or an annulus to a radial Gaussian confined to a thin annulus
\end{abstract}

\maketitle

\section{Introduction}

One of the most striking phenomena in condensed matter physics is the Fractional Quantum Hall Effect (FQHE) for charged fermions in strong magnetic fields that still, after decades of research, poses many challenging questions \cite{STG,Gir,Goe}.

The formal similarity between the Hamiltonian of a rotating Bose gas and that of a 2D electron gas in a magnetic field suggests the possibility of studying {\em bosonic} analogues of the FQHE in cold quantum gases set in rapid rotation \cite{WG,VHR, CWG, Co,Vie}. In particular, strongly correlated phases analogous to the fermionic Laughlin state \cite{Lau, Lau2} are predicted to occur in rotating boson clouds. More specifically, a phase transition from a Bose-Einstein condensate (BEC) with a vortex lattice to a strongly correlated state should happen at total angular momentum $\propto N ^2$ (filling factor of order $1$), where $N\gg 1$ denotes the particle number.  This phase transition has not been observed yet because it is extremely difficult to reach such high angular momenta in the laboratory.

The reasons for this difficulty can be understood as follows. The usual experimental setup consists of bosonic atoms in a rotating trap modeled by a harmonic potential. Once the usual approximations have been made, the relevant Hamiltonian boils down to 
\begin{equation}\label{eq:LLLh harm'}
H=  \sum_{j=1} ^N  \om |x_j| ^2+ g \sum_{i<j} \delta (x_i-x_j)
\end{equation}
operating on wave functions $\Psi(x_1,\dots, x_N)$, $x_i\in\mathbb R^2$, in the  lowest Landau level (LLL) of a two-dimensional magnetic Hamiltonian. Here $\om>0$ is half the difference between the squares of the frequency of the harmonic trap and of the rotation velocity, measured in units of the latter, and $g\geq 0$ is the coupling constant of the interaction. To obtain correlated states, one wants to favor the interaction, i.e., consider small values of $\om/g$. Clearly this corresponds to a singular limit where the gas is no longer confined against centrifugal forces. This is the main source of  difficulty that has to be overcome to create the Bose-Laughlin state (see \cite{RRD} for a more quantitative discussion and \cite{BSSD,SCEMC} for state of the art experiments in this direction). 

A natural proposal \cite{BSSD,Vie,MF} is to add a confining term that should be stronger than quadratic (i.e., stronger than the centrifugal force). In the present paper we investigate asymptotic properties of ground states after such an addition to the Hamiltonian  for large particle numbers and when the parameters are tuned to favor strong correlations. A simple model for the additional term is $\sum_j k|x_j|^4$ with $k>0$, that has already been considered in the context of rotating BECs \cite{CPRY,FJS}. Besides the quest for the Laughlin state, our motivation is also to explore new phenomena arising from this modification of \eqref{eq:LLLh harm'}. 

A new problem that has to be tackled is the lack of commutativity of the anharmonic potential term, projected onto the lowest Landau level,  and the interaction. This implies that the Laughlin state is not an exact eigenstate of the modified Hamiltonian as it is for the Hamiltonian in a purely harmonic trap. Nevertheless we prove that Laughlin-like correlations can be achieved asymptotically by a suitable choice of the parameters $\omega$ and $k$ in dependence of $N$. Moreover, we argue that the ground state exhibits a {\it new strongly correlated phase} if  $\omega$ is negative and the ratio $|\omega|/k $ of order $N$ or larger. In this phase the density is depleted in a hole around the center of the trap while it has the features of an incompressible quantum fluid in an annulus around the hole. In fact, there are clear indications of a {\it further} transition at 
$|\omega|/k \gtrsim N^2$ with the density concentrated in a thin annulus of large radius with a Gaussian profile in the radial variable, while for $|\omega|/k $ significantly smaller that $N^2$ the profile in the annulus is essentially flat.

Our analysis relies on the one hand on energy estimates, making use of a representation of the anharmonic term through angular momentum operators, and on the other hand on quantitative estimates for the one-particle density of strongly correlated trial states. These states have the form of a Laughlin state modified by a factor corresponding to a vortex of high angular momentum at the origin  (`Laughlin times giant vortex' or Laughlin quasi-hole). Mathematically the energy estimates are much simpler than the second part of the analysis and already lead to criteria for strong correlations in an asymptotic limit and estimates on the angular momentum of the ground state. The estimates on the density, on the other hand, are crucial for a physical interpretation of the two phase transitions indicated by the energy estimates. Moreover, the density estimates imply an improvement of the energy estimates. 

The intuition behind the density estimates is Laughlin's plasma analogy \cite{Lau, Lau2}: The trial states are seen as defining Gibbs measures of a 2D classical Coulomb gas.  The one-body densities appear as combined mean field/low temperature limits for the Coulomb gas of the type studied e.g.\ in \cite{MS,Kie,CLMP,KS} using compactness arguments. For quantitative error estimates, a new method for the study of the mean-field limit is needed. In the present paper we outline this method but refer to a companion paper \cite{RSY} for full details.

The depletion of the density at the center of the trap and absence of vortices in the bulk for appropriate values of the parameters is a feature common to the so-called `giant vortex' phase of {\it uncorrelated} states in the Gross-Pitaevskii (GP) approximation. The transition from a vortex lattice to a giant vortex phase in anharmonic traps has been studied by various methods over the past 10 years, e.g., in \cite{Fe2, KTU, FB, KB, AD, FZ}, and was recently established with full mathematical rigor (see \cite{CPRY} for references). We stress, however, that this transition is physically and mathematically quite different from the one studied here.

The rest of the paper is organized as follows: In Section \ref{sec:main results} below we state our results on the energy estimates and their consequences  in the form of three theorems. The energy upper and lower bounds match to leading order, and  the change of optimal trial functions  when the parameters are varied  indicates a transition from one type of strongly correlated ground state (Laughlin type) to another (Laughlin times giant vortex). The precise statements of the results requires some preliminary discussions in Section \ref{sec:hamil} of the properties of the Hamiltonian \eqref{eq:LLLh harm'} regarded as an operator on a Bargmann space of analytic functions. In Section \ref{sec:anharmonic} we introduce the anharmonic addition to the potential as well as its alternate definition through the square of the one-particle angular momentum operator in the lowest Landau level. The theorems stated in Section \ref{sec:main results} are proved in Section \ref{sec:proofs}. In Section \ref{sec:plasma} we 
discuss the determination of the one-body density in the strongly correlated trial states considered and its implications for the interpretation and improvement of the results of Section \ref{sec:main results}. A final Section \ref{sec:conclu} summarizes our findings and discusses their relation to the recent paper \cite{RRD}.

More details on our approach can be found in the companion paper \cite{RSY}.
 
\section{The standard Hamiltonian}\label{sec:hamil}

It is well known \cite{GJ,PB} (see also \cite{LSY,LS} for more references) that the Hamiltonian \eqref{eq:LLLh harm'} in the LLL can be regarded as an operator on the Bargmann space $\mathcal B_N$ of symmetric {analytic} functions $ F$ of $z_1,\dots,z_N$ with $z_i\in \mathbb C$ such that
\beq \label{scalarprod}
\int_{\mathbb C^N}| F(z_1,\ldots,z_N)|^2\exp\big(-\hbox{$\sum_{j=1}^N$}|z_j|^2\big)\,\mathrm dz_1\cdots \,\mathrm dz_N<\infty. 
\eeq 
Here $\mathrm d z$ denotes the Lebesgue measure on $\mathbb R^2$ that is identified with the complex plane $\mathbb C$ in the usual way. On $\mathcal B_N$ the angular momentum operator of the $i$-th particle is $L_i=z_i\partial _{z_i}$ and the contact interaction $\delta(z_i-z_j)$ is given by the bounded operator
\begin{multline} 
\delta_{ij} F(\ldots,z_i,\ldots,z_j\dots)\\=\frac 1{2\pi} F\big(\ldots,\half(z_i+z_j), \ldots,\half(z_i+z_j),\ldots\big)
\end{multline} 
as noted first in \cite{PB}. Indeed, using the analyticity of $ F$, it is easy to see that
{\begin{multline} 
\langle  F,\delta_{ij} F\rangle=\int_{\mathbb C^N}|{ F(\ldots,z,\ldots,z,\ldots)}|^2\exp(-2|z|^2)\,\mathrm dz\\ \times \exp(-\hbox{$\sum_{k\neq i,j}$}|z_k|^2)\hbox{$\prod_{k\neq i,j}$}dz_k,
\end{multline}
where $\langle\cdot\rangle$ is the scalar product on $\mathcal B_N$ corresponding to \eqref{scalarprod}. Using that $z\partial_z\leftrightarrow |z|^2-1$ in this scalar product the operator \eqref{eq:LLLh harm'} can, apart from a constant additive term $N\omega$, be written as
\beq\label{yrastham} 
H_N= {\omega}\,\mathcal L_N+{g}\,\mathcal I_N
\eeq  
with 
\beq
\mathcal L_N=\sum_{i=1}^Nz_i\partial_{z_i}\, \quad\quad \mathcal I_N=\sum_{i<j}\delta_{ij}.
\eeq 
We denote by $\mathcal H_N$ the space of wave functions of the form $\Psi(z_1,\dots,z_N)= F(z_1,\dots,z_N)\exp(-\sum_{j=1}^N|z_j|^2/2)$ with $ F\in \mathcal B_N$. This is a subspace of $L^2(\mathbb R^{2N})$ and clearly isomorphic to $\mathcal B_N$. Note, however, that the angular momentum operator $L_i$ on $\mathcal H_N$ is $z_i\partial_{z_i}-\bar z_i\partial_{\bar z_i}$ rather than $z_i\partial_{z_i}$ as on $\mathcal B_N$. To distinguish these two realizations of states in the LLL we shall denote functions in $\mathcal B_N$ by roman letters and in $\mathcal H_N$ by greek letters.

We remark that the reduced Hamiltonian \eqref{yrastham} is obtained by restricting states to the LLL associated with the rotational frequency (that we fix equal to $1$), which leads to \eqref{eq:LLLh harm'}, and then using the substitution $z\partial_z\leftrightarrow |z|^2-1$. Another possibility (see \cite{LS,LSY,Vie} and references therein) is to restrict states to the LLL associated with the harmonic trap frequency, which directly leads to the model \eqref{yrastham} but with $\om$ equal to the difference between the two frequencies (rather than half the difference between the squares). Since in the regime of interest the two frequencies are close, the two approaches are equivalent. Ours has the advantage of better emphasizing what the residual effective potential is, which is natural when we consider the additional anharmonic term below. 

An essential feature of the Hamiltonian \eqref{yrastham}  is that the operators $\mathcal L_N$ and{ $\mathcal I_N$ {commute}. The  lower boundary of (the convex hull of) their joint spectrum in a plot with angular momentum as the horizontal axis is called the  {\it yrast curve} (see, e.g., \cite{Vie} and \cite{LS} for plots showing its qualitative features). As a function of the eigenvalues $L$ of $\mathcal L_N$ the yrast curve $I(L)$ is decreasing from $I(0)= (4\pi)^{-1} N(N-1)$ to $I(N(N-1))=0$. The monotonicity follows from the observation that if a simultaneous eigenfunction of $\mathcal L_N$ and $\mathcal I_N$ is multiplied by the center of mass, $(z_1+\cdots+z_N)/N$, the interaction is unchanged while the angular momentum increases by one unit. 

For a given ratio $\omega/g$ the ground state of \eqref{yrastham} (in general not unique) is determined by the point(s) on the yrast curve  where a supporting line has slope $-\omega/g$.

\subsection{Fully correlated states}\label{sec:correl states}

For $L\leq N$ the ground state of \eqref{yrastham} is explicitly known (see \cite{SW} or \cite{Vie} and references cited there) while for large $N$ and $L\ll N^2$ a Gross-Pitaevskii description with an uncorrelated ground state is asymptotically correct \cite{LSY}. Results about the Gross-Pitaevskii functional with states restricted to the lowest Landau level may be found in \cite{ABN1,ABN2,BR,R}.

For $L=N(N-1)$ the unique ground state of $\mathcal I_N$ with eigenvalue $0$ is the bosonic {\it  Laughlin state}  whose wave function
in $\mathcal B_N$ is the symmetric polynomial
\beq\label{Lau} F_{\rm Lau}(z_1,\dots,z_N)={\rm const.}\prod_{i<j}(z_i-z_j)^2\eeq
or, equivalently, in $\mathcal H_N$
\beq\Psi_{\rm Lau}(z_1,\dots,z_N)={\rm const.} \prod_{i<j}(z_i-z_j)^2\exp(-\hbox{$\sum_k$}|z_k|^2/2).\eeq
More generally, the null space ${\rm Ker}\,\mathcal I_N$ of the interaction operator consists of functions of the form \beq\label{fullycorrelated} F(z_1,\dots, z_N)=G(z_1,\dots, z_N) F_{\rm Lau}(z_1,\dots, z_N)\eeq where $G$ is a symmetric analytic function such that $F$ is square integrable w.r.t.\ $\prod_k\exp(-|z_k|^2)\mathrm dz_k$. We shall call states of this form {\it fully correlated states}. Their angular momentum spectrum is contained in $L\geq N(N-1)$.

A common interpretation of the form \eqref{Lau} is that each particle ``binds'' a vortex a degree $2$ (see e.g. \cite{STG,Vie} and references therein). Indeed, since LLL wave functions are analytic, putting a zero at each particle in order to cancel the interaction energy requires a non trivial quantized phase. 

\subsection{Spectral gaps}\label{sec:gaps}

States with a given eigenvalue $L$ of the total angular momentum correspond to symmetric, homogeneous polynomials of degree $L$. This is a finite dimensional space and for all $L$ it contains states with strictly positive interaction energy. (Take for example $ F=({\rm const.})\sum_k z_k^L$.)  Hence the spectral gap
\beq \Delta(L)\equiv\text{smallest eigenvalue $>0$ of}\ \mathcal I_N\upharpoonright_{\{\mathcal L_N=L\}}\eeq
is strictly positive for all $L$. Note that $\Delta(L)$ depends on $N$ besides $L$.

The trial state argument proving the monotonic decrease of the yrast curve also applies to $\Delta(L)$, and proves that it is decreasing with $L$. Numerical studies \cite{RCJJ,RJ1,RJ2,VHR} suggest that $\Delta\equiv \inf_L \Delta(L)=\Delta(N(N-1)-1)>0$ and it is also believed (see the discussion in \cite{LS}) that $\Delta$ stays of order 1 and bounded away from zero as $N\to\infty$. If this {\it spectral gap conjecture} is true, the statements of Theorems \ref{teo:correl} and \ref{teo:ener}  below can be simplified, but to our knowledge the conjecture has not yet been proved. We shall therefore not rely on any such assumptions but state our results in the sequel in terms of $\Delta(L)$.

According to numerical diagonalizations \cite{RJ1,RJ2,VHR}, it seems that the yrast state corresponding to the interaction energy $\Delta(N(N-1)-1)$ is a center of mass excitation of the the state corresponding to $\Delta (N(N-1)-N)$, which would imply $\Delta(N(N-1)-1) = \Delta(N(N-1)-N)$. The latter quantity is associated with the energy cost to create a Laughlin quasi-particle \cite{Lau,Lau2} in the system. We do not know a proof that this picture is correct, however. 


\section{Adding an anharmonic potential}\label{sec:anharmonic}

The limit $\omega\to 0$ {keeping $\omega>0$ is experimentally very delicate. For  stability, but also to study new effects, we consider now a modification of the Hamiltonian where one adds a quartic term to the potential:
\beq 
H \rightarrow H +k\sum_{j=1}^N |z_j|^4
\eeq
with a {new parameter} $k>0$.  The expectation value of the energy for a normalized $\Psi=F\cdot\exp(-\sum_i |z_i|^2/2) \in\mathcal H_N$  is given by the functional
\beq\label{expvalue} 
\mathcal E[\Psi]=\int V_{\omega,k}(z)\rho_\Psi(z)+g\langle F,\mathcal I_N F\rangle
\eeq 
where 
\beq 
V_{\omega,k}(z)=\omega\, |z|^2 + k\,|z|^4
\eeq 
and $\rho_\Psi$ is the one-body density with the  normalization 
\beq \label{poten}\int \rho_\Psi(z)\,\mathrm dz=N.\eeq
In particular, the energy is bounded below by $N\min V_{\omega,k}$ that is finite even if $\omega<0$, provided $k>0$. 
For the fully correlated states \eqref{fullycorrelated} the second term in \eqref{expvalue} vanishes and the expectation value of the energy is entirely determined by the one-particle density.

Using the correspondence $z\partial_z\leftrightarrow |z|^2-1$ and $(z\partial_z)^2\leftrightarrow|z|^4-3|z|^2+1$ the modified Hamiltonian on the Bargmann space $\mathcal B_N$ can be written (up to the additive constant $N(\omega+2k)$ that will be dropped in the following) as
\beq\label{hprime} 
H_N'=(\omega+{3k})\mathcal L_N+{k} \sum_{i=1}^N L_i^2+g\,\mathcal I_N.
\eeq 
For the original Hamiltonian \eqref{yrastham}, without the anharmonic addition to the potential, the Laughlin state is an {\it exact} eigenstate  with $L=N(N-1)$ and energy $\omega N(N-1)$. For $k\neq 0$, however, the Laughlin state is not an eigenstate of $H_N'$ because
$ \sum_{i=1}^N L_i^2$ {does not commute} with $\mathcal I_N$. On the other hand $\mathcal L_N$ still commutes with $H_N'$ and we may consider the ground state energy $E_0(L)$ of $H_N'$ restricted to the subspace where $\mathcal L_N=L$. The unrestricted ground state energy $\min_L E_0(L)$ will be denoted by $E_0$, a ground state (that might not be unique) by $\Psi_0$ and its angular momentum by $L_0$.

\section{Asymptotic properties of ground states}\label{sec:main results}

We now state our main results concerning  the angular momentum $L_0$, the confinement to the space of fully correlated states, and the energy of a ground state $\Psi_0$  of \eqref{hprime} in the limit when $N\to\infty$ and at the same time $\omega,\, k\to 0$.
Note that $\Psi_0$ can be assumed to have a definite angular momentum because $\mathcal L_N$ commutes with \eqref{hprime}.

\begin{teo}[\textbf{Angular momentum}]\label{teo:mom}\mbox{}\\
 In the limit $N\to \infty$, $\om,k \to 0$ the angular momentum $L_0$ of a ground state of $H_N'$ satisfies
\begin{itemize}
\item If $\om \geq - 2 k N$
\begin{equation}\label{16}
L_0 \leq 2 N ^2 
\end{equation}
\item If $\om \leq -2 kN$  but $|\omega|/k\lesssim N^2$
{\begin{equation}
\label{17}\left| L_0 - L_{\rm qh} \right| \leq \sqrt 3\, N ^2 (1+o(1)),
\end{equation}}
where
\begin{equation}
L_{\rm qh} :=  \frac{|\om| N}{2k}.
\end{equation}
\item If $\om \leq -2 kN$ and $|\omega|/k\gg N^2$
\beq\label{19}
\left| L_0 - L_{\rm qh} \right| \leq \sqrt 3\, L_{\rm qh}^{1/2} N (1+o(1)). 
\eeq
\end{itemize}
In particular,  $L_0/L_{\rm qh}\to 1$ in both cases \eqref{17} and \eqref{19}.
\end{teo}

The next theorem provides criteria for a ground state to be asymptotically fully correlated, i.e., in the kernel of the interaction operator. In the statement the following gaps in the spectrum enter:
\begin{multline} \Delta_1\equiv \Delta (2N^2),\quad \Delta_3\equiv\Delta(L_{\rm qh}+ \sqrt 3(N^{2})),\\ \Delta_4\equiv \Delta(L_{\rm qh}+ \sqrt 3(L_{\rm qh}^{1/2}N)).\end{multline}
If the spectral gap conjecture holds, all these gaps can be replaced by the minimal gap $\Delta>0$ independent of $N$ and the parameters.

\begin{teo}[\textbf{Criteria for full correlation}]\label{teo:correl}\mbox{}\\
Let $\PKerp$ denote the projector on the orthogonal complement of the kernel of $\mathcal I_N$. We have
{\begin{equation}
\left\Vert \PKerp \Psi_0 \right\Vert_{L^2 (\R ^{2N})}  \to 0  
\end{equation}}
in the limit $N\to \infty$, $\om,k\to 0$ if one of the following conditions holds:\smallskip

\noindent {\bf Case 1.} $\om \geq 0$ and
$(g\,\Delta_1)^{-1}\cdot(\omega N^2+kN^3)\rightarrow 0.$
In this case, for $N$ large enough, 
\beq \label{26}\Vert \PKerp \Psi_0\Vert^2\leq (g\,\Delta_1)^{-1}\cdot 3 kN^3(1+o(1)).\eeq

\noindent {\bf Case 2.} $- 2 k N\leq \om \leq 0$ and
$ (g\,\Delta_1)^{-1}\cdot({N (\om ^2}/{k}) + \om N ^2 + k N ^3) \rightarrow 0
$. Again \eqref{26} holds under this condition for $N$ large.\smallskip

\noindent{\bf Case 3.} $\om \leq - 2 k N$, $|\omega|/k\lesssim N^2$ and
$
 (g\,\Delta_3)^{-1} \cdot k N ^{3}\rightarrow 0.
$
Here 
\beq
\Vert \PKerp \Psi_0\Vert^2\leq   (g\,\Delta_3)^{-1}\cdot  \left(3k\,N^3+\frac 32|\omega|N\right)(1+o(1)).
\eeq 
\noindent{\bf Case 4.}
$\om \leq - 2 k N$, $|\omega|/k\gg N^2$ and
$
 (g \, \Delta_4)^{-1}\cdot |\omega|N \rightarrow 0.
$
Here 
\beq\Vert \PKerp \Psi_0\Vert^2\leq (g\,\Delta_4)^{-1} \cdot\frac 32 |\omega| N\,  (1+o(1)).\eeq 
\end{teo}

Both Theorems 1 and 2 are consequences of the energy bounds stated in the following Theorem 3. The ``cases" referred to are the same as in Theorem 2.
\begin{teo}[\textbf{Energy bounds}]\label{teo:ener}\mbox{}\\ 
The ground state energy $E_0$ satisfies the following bounds:\medskip

\noindent {\bf Cases 1 and 2}
\beq\label{25} (\omega N^2+kN^3)(1-o(1))\leq E_0\leq \omega N^2+4 kN^3.\eeq
\noindent {\bf Case 3}
\beq -\frac {\omega^2N}{4k}(1+o(1))\leq E_0\leq -\frac {\omega^2N}{4k}+ \left(3k N^3+\frac 32|\omega|N\right)(1+o(1)).\eeq
\noindent {\bf Case 4}
\beq -\frac {\omega^2N}{4k}(1+o(1))\leq E_0\leq -\frac {\omega^2N}{4k}+\frac 32 |\omega| N(1+o(1)).\eeq
\end{teo}

Note that the present paper is only concerned with properties of the ground states of the system. The usual (heuristic and numerical) arguments supporting the picture of fractionally charged excitations of the ground states \cite{Lau,Gir} carry over with little modifications in our setting. They are not on the same level of mathematical rigor as the above theorems however, and we have no rigorous arguments allowing to identify the excited states of our model.

\section{Proofs of Theorems 1-3}\label{sec:proofs}
The proofs of Theorems 1--3 rely on two ingredients:
\begin{itemize}
\item A lower bound for the energy at fixed angular momentum $L$:
\beq \label{lowerbound} E_0(L)\geq (\omega+3k)L+k\frac {L^2} N.\eeq
\item An upper bound for the energy of suitable trial functions.
\end{itemize}

The proof of \eqref{lowerbound} is quite simple: Dropping the nonnegative interaction term in \eqref{hprime} one has to minimize $\langle F, ((\omega+{3k})\mathcal L_N+{k} \sum_{i=1}^N L_i^2) F\rangle$ over $ F\in\mathcal B_N$ with $\mathcal L_N F=L F$. The bound follows from the operator inequality
$\sum_i L_i^2\geq \frac 1N(\sum_i L_i)^2$ that holds because $L_i$ and $L_j$ commute for any $i,j$.

The upper bounds on $E_0$  can be derived as follows. As trial functions in the Bargmann space $\mathcal B_N$ 
we take $F_{\rm qh}^{(m)}\equiv GF_{\rm Lau}$ with $F_{\rm Lau}$ as in \eqref{Lau} and
\beq \label{eq:trial states}
G(z_1,\dots, z_N)=c_{N,m}\prod_{i=1}^Nz_i^m
\eeq
with $c_{N,m}$ such that $F_{\rm qh}^{(m)}$ is normalized. One recovers the pure Laughlin state for $m=0$ and the states with $m>0$ are often referred to as `Laughlin quasi-holes'. They already appeared in Laughlin's seminal paper \cite{Lau}. Note that we will take a large value of $m$ in the sequel, in which case the denomination `Laughlin times giant vortex' seems more appropriate.

Since the trial states \eqref{eq:trial states} all have a zero interaction energy, we only have to estimate their potential energy. We have
\begin{multline}\label{30} 
\mathcal L_N F_{\rm qh}^{(m)}=(\mathcal L_N G)F_{\rm Lau}+G(\mathcal L_NF_{\rm Lau})=\\
(Nm+N(N-1)) F_{\rm qh}^{(m)}.
\end{multline}
Moreover,
\begin{multline}\label{31}\sum_{i=1}^N\langle F_{\rm qh}^{(m)}, L_i^2F_{\rm qh}^{(m)}\rangle=
\sum_{i=1}^N\langle L_iF_{\rm qh}^{(m)}, L_iF_{\rm qh}^{(m)}\rangle=\\ Nm^2 + 2m N(N-1)+N\langle G L_1 F_{\rm Lau}, G L_1 F_{\rm Lau}\rangle
\end{multline}
where the symmetry of $F_{\rm Lau}$ has been used for the last term which can be written as
\beq 
N\int |L_1F_{\rm Lau}|^2|G|^2\exp(-\sum_i|z_i|^2)\prod_i\mathrm dz_i.
\eeq
We now note that the monomials $z_1^n$ are orthogonal for different $n$'s in the $L^2$-scalar product $\langle .\:, \: . \rangle_G $ on analytic functions defined by the measure $|G|^2\exp(-\sum_i|z_i|^2)\prod_i\mathrm dz_i$  because $|G|^2$ depends only on the absolute values of the $z_i$'s. Moreover, $L_1z_1^n=nz_1^n$, and $F_{\rm Lau}$ is a polynomial of degree $2(N-1)$ in $z_1$. Hence
\begin{align} \label{33}
N\langle G L_1 F_{\rm Lau}, G L_1 F_{\rm Lau}\rangle&\leq 4N(N-1)^2 \langle F_{\rm Lau},  F_{\rm Lau}\rangle_G \nonumber \\
&= 4N(N-1)^2 \langle G F_{\rm Lau},  G F_{\rm Lau} \rangle \nonumber \\
&=  4N(N-1)^2
\end{align}
where we has also used the normalization of our trial state. Inserting Eqs.\ \eqref{30}, \eqref{31} and \eqref{33} in the expectation value $\langle F_{\rm qh} ^{(m)},H_N'F_{\rm qh}  ^{(m)} \rangle$ leads to the upper bound 
\begin{multline}
\langle F_{\rm qh} ^{(m)},H_N'F_{\rm qh}  ^{(m)} \rangle \leq (\om + 3k)(mN + N(N-1)) \\
+ k (Nm ^2 + 2 N(N-1)m + 4 N ^3). 
\end{multline}
Optimizing over $m$ gives the value 
\begin{equation}\label{eq:intro m opt}
\mopt = \begin{cases}
               0 \mbox{ if } \om \geq - 2 k N \\
               \frac{|\om|}{2 k}- N \mbox{ if } \om < - 2 k N.
              \end{cases} 
\end{equation}
Strictly speaking $\mopt$ must be an integer so we should take the integer value of the above. Since we are only concerned with orders of magnitudes this does not produce any change in the final result, and we ignore this detail. 
For $\mopt=0$ this gives the upper bound in \eqref{25}. 

For $\mopt=({|\om|}/{2 k})- N>0$ we obtain
\beq\label{35} E_0\leq -\frac {\omega^2N}{4k}+ \left(3 kN^3+\frac 32 |\om|N \right)(1+o(1)).\eeq 
The first term in \eqref{35} is equal to $N$ times the minimum of the potential $\omega|z|^2+k|z|^4$. The second term $\sim kN^3$ dominates the third term $\sim|\omega|N\sim kN\,\mopt$ for $ |\omega|/2k\ll N^2$, while the converse holds for $ |\omega|/2k\gg N^2$. Note also that $ L_{\rm qh}\approx N \mopt$ for $|\om|/2k\gg N$. 

To proceed we note that by \eqref{lowerbound} and the definition of the gap we have
\beq\label{36} (\omega+3k)L_0+k\frac {L_0^2}N+g\,\Delta (L_0)\Vert \PKerp\Vert^2\leq E_0.
\eeq

For $\omega\geq -2kN$ it follows from  \eqref{36}  and the upper bound in \eqref{25} that 
\beq \label{37}(\omega+3k)L_0+k\frac {L_0^2}N\leq \omega N^2+4 kN^3\eeq
which implies
\beq\label{38}{L_0\leq 2N^2}\eeq
and hence
\beq\label{39} \Delta(L_0)\geq \Delta(2N^2).\eeq
For $\omega\geq 0$, $f(L)\equiv (\omega+3k)L+k{L^2}/N$ is nonnegative. Hence we see that in this case
$\Vert \PKerp \Psi_0\Vert^2\rightarrow 0$ if 
\beq\label{40}{
\{g\,\Delta (2N^{2})\}^{-1}(kN^3+\omega N^2)\rightarrow 0.}
\eeq
On the other hand, if $\Vert \PKerp \Psi_0\Vert^2\rightarrow 0$, then it is clear that $L_0$ must be $\geq N(N-1)$ for $N$ large enough, and hence $f(L_0)\geq (\omega N^2+kN^3)(1-o(1))$. 

Altogether we have, for $\omega\geq 0$ and under the condition \eqref{40}, the bounds
\beq\label{41} {}{(\omega N^2+kN^3)(1-o(1))\leq E_0\leq \omega N^2+4 kN^3(1+o(1))}\eeq
as well as
\beq\label{42} {}{\Vert \PKerp \Psi_0\Vert^2\leq 3 \{g\,\Delta (2N^{2})\}^{-1} kN^3 (1+o(1)).}\eeq 

Consider next the case $-2kN\leq \omega\leq 0$. As before, Eq.\ \eqref{37} implies \eqref{38} and \eqref{39}.
For the analogue of \eqref{40} we use that
$f(L)$ takes a minimal value at 
\beq L_{\rm min}=-\frac N {2k}(\omega+3k)=L_{\rm qh}(1+o(1))\eeq
with
\beq\label{44} f(L_{\rm min})=-\frac N {4k}(\omega+3k)^2=-\frac{\omega^2 N}{4k}(1+o(1)).\eeq
Condition \eqref{40} for a fully correlated ground state  thus gets replaced by 
\beq\label{45}{}{
\{g\,\Delta (2N^{2})\}^{-1}\left(kN^3+\omega N^2+N\frac{\omega^2}k\right)\rightarrow 0.}
\eeq
Moreover, \eqref{45} implies as before that $L_0$ eventually becomes $\geq N(N-1)$. For $-2kN\leq \omega\leq 0$, 
$L_{\rm min}$ can take values between $0$ and $N(N-\hbox{$\frac 32$})$ so we can conclude that $f(L_0)\geq
f(N(N-1))$ that leads again to \eqref{41} and \eqref{42}.

For the case $\omega\leq -2kN$ we use
\beq f(L)= f(L_{\rm min})+\frac kN(L-L_{\rm min})^2\eeq
and obtain from \eqref{36} and \eqref{44}, up to factors $(1+o(1))$,
\beq\label{1.15}\frac kN(L_0-L_{\rm min})^2+g\,\Delta(L_0)\Vert \PKerp \Psi_0\Vert^2\leq 3 kN^3.\eeq
In particular,
\beq\label{1.16}{}{|L_0-L_{\rm min}|\leq {\sqrt 3} N^2}\eeq
and thus $\Delta(L_0)\geq \Delta(L_{\rm min}+\sqrt 3N^2)$. The condition for full correlation, analogous to \eqref{40} and \eqref{45} becomes
\beq\label{1.17}\{g\,\Delta(L_{\rm min}+\sqrt 3N^2)\}^{-1}k\,N^3\rightarrow 0,\eeq
indeed,  
\beq\label{1.18} {}{\Vert \PKerp \Psi_0\Vert^2\leq \frac 13 \{g\,\Delta(L_{\rm min}+\sqrt 3N^2)\}^{-1}k\,N^3}.\eeq
Since $f(L_{\rm min})=-N\omega^2/4k$ to leading order, the energy estimate is
\beq\label{1.19} {}{-\frac {\omega^2N}{4k}(1-o(1))\leq E_0\leq -\frac {\omega^2N}{4k}+ 3 kN^3(1+o(1)).}\eeq
The condition for $kN^3\ll \omega^2N/k$ is equivalent to $|\omega|/k\gg N$, i.e., $m_{\rm opt}\gg N$. Note also that \eqref{1.16} implies
\beq\label{1.20}  \frac{|L_0-L_{\rm qh}|}{L_{\rm qh}}\ll 1\eeq
if $m_{\rm opt}\gg N.$

If  $- \omega/{k}\gg N^2$ as in Case 4,  the error term $3 kN^3$ in \eqref{35} is small compared to  $\frac 32|\omega| N\sim kN\,\mopt$ and we have  
\beq\label{} E_0\leq -\frac {\omega^2N}{4k}+ \frac 32 |\om|N (1+o(1)).\eeq 
Thus \eqref{1.15} gets replaced by (as always, up to $(1+o(1)$)
\beq\label{1.21}\frac kN(L_0-L_{\rm min})^2+g\,\Delta(L_0)\Vert \PKerp \Psi_0\Vert^2\leq \frac 32| \omega| N.\eeq
and \eqref{1.16} by
\beq\label{1.22}{}{|L_0-L_{\rm min}|\leq {\sqrt 3} \left(\frac {|\omega|}{2k}\right)^{1/2} N}\eeq
which  implies \eqref{19} and \eqref{1.20} for $m_{\rm opt}\gg N^2$.

The analogues of \eqref{1.17}, \eqref{1.18} and \eqref{1.19} for $m_{\rm opt}\gg N^2$ are
\beq\{g\,\Delta(L_{\rm min}+\sqrt 3NL_{\rm min}^{1/2})\}^{-1}|\omega|N\rightarrow 0,\eeq

\beq {}{\Vert \PKerp \Psi_0\Vert^2\leq \frac 32 \{g\,\Delta(L_{\rm min}+\sqrt 3NL_{\rm min}^{1/2})\}^{-1}|\omega| N, }\eeq
and
\beq {}{-\frac {\omega^2N}{4k}(1-o(1))\leq E_0\leq -\frac {\omega^2N}{4k}+\frac 32 |\omega| N(1+o(1)).}\eeq


\section{The one-particle density of fully correlated trial states}\label{sec:plasma}

The main conclusions that can be drawn from the previous sections may be summarized as follows:
\begin{itemize}
\item The energy estimates are consistent with the picture that the Laughlin state is an approximate ground state for positive $\om$ in Case 1,  and also for  negative $\omega$ in Case 2, as long as $|\omega|/k\lesssim N$. The angular momentum remains $O(N^2)$ in these cases.
\item When $\omega<0$ and  $|\omega|/kN$ becomes large (Cases 3 and 4) the angular momentum is approximately  $L_{\rm qh}=O(N|\omega|/k)\gg N^2$, much larger than for the Laughlin state.
\item A transition between Cases 3 and 4 at $|\omega|/k\sim N^2$ is manifest through the change of the subleading contribution to the energy of the trial functions. Its order of magnitude changes from $O(kN ^3)$ to $O(|\om| N)$ at the transition.
\end{itemize}

We now want to substantiate this picture further by investigating the one-particle densities of our trial states, This will both clarify the physical meaning of the two transitions at $|\omega|/k\sim N$ and
$|\omega|/k\sim N^2$ that are indicated by the energy considerations above, and also allow improvements of the energy estimates. The latter rely on the representation \eqref{expvalue} of the energy and requires detailed estimates of the one-particle density. We note that the representation  \eqref{expvalue} holds also for more general potentials than $\omega|z|^2+k|z|^4$ (e.g., anisotropic potentials), where the approach via angular momentum presented in Section V would not apply. We shall here present the main results of the analysis of the density but refer to \cite{RSY} for details of the proofs.

Already Laughlin's  paper  of 1983 \cite{Lau} contains the important idea that the determination of the probability density $|\Psi|^2$ in his and related states for large $N$ is equivalent to a {2D electrostatic problem}.
We make use of this idea for our trial states  whose wave functions in $\mathcal H_N$ are
\beq \label{eq:quasi-hole} 
\Psi_{\rm qh}^{(m)}(z_1,\dots,z_N) = c_{N,m} \prod_{j=1} ^N z_j ^m \prod_{i<j} \left( z_i - z_j\right) ^2 e^{-\sum_{j=1} ^N |z_j| ^2 / 2} 
\eeq  
where $c_{N,m}$ is a normalization constant. We denote $(z_1,\dots,z_N)$ by $Z$ for short and consider the scaled probability density
\beq\label{scaledmu} 
\muNm (Z) \equiv N ^N \left| \Psi_{\rm qh}^{(m)} (\sqrt{N} Z )\right| ^2.
\eeq 
Then we can write
{\begin{align}
\muNm (Z) &= \mathcal Z_{N,m}^{-1} \exp\left\{ \sum_{j=1} ^N \left( - N  |z_j| ^2 + 2 m \log |z_j|\right)\right. \nonumber\\ 
& \left.  - 4 \sum_{i<j} \log |z_i - z_j|\right\} \nonumber \\
&= \mathcal Z_{N,m}^{-1} \exp\left( -\frac{1}{T} \HNm(Z) \right)
\end{align}}
with $T=N^{-1}$ and 
\beq\label{hamfct} 
\HNm(Z)=\sum_{j=1} ^N \left( |z_j| ^2 - \frac{2 m}N \log |z_j|\right)  - \frac 4N \sum_{i<j} \log |z_i - z_j|.
\eeq 
The normalization factor (partition function) is 
$$\ZNm=\int \exp\left( -\frac{1}{T} \HNm(Z) \right)\,\mathrm d^{2N}Z.$$
In the sequel we discuss the  properties of the density $\muNm (Z)$ in the 
$N\to\infty$ limit.

\subsection{Mean field limit}\label{sec:mean field}

The Hamiltonian \eqref{hamfct} describes a classical 2D Coulomb gas in a uniform  background of opposite charge and a point charge of strength $2m/N$ at the origin, corresponding respectively to the $|z_j|^2$ and the $-\frac{2 m}N \log |z_j|$ terms. The probability measure $\muNm (Z)$ minimizes the free energy $F(T,N,m)=-T\log \ZNm$ for this Hamiltonian at $T=N^{-1}$. 

The $N\to\infty$ limit is in this interpretation a mean field limit at small temperature $T\to 0$.  Thus, it is reasonable to expect that in the limit $\muNm$ factorizes, $\muNm\approx \rho^{\otimes N}$, where the probability measure $\rho$ on $\mathbb R^2$ minimizes an appropriate mean field energy functional. In fact, using compactness arguments, rigorous results of this type have been obtained in \cite{CLMP,Kie,KS} for related models. The computation of the energy via Equation \eqref{expvalue}, however, requires quantitative estimates of the errors in the approximations that can only be obtained by a different method. A more elaborate analysis is developed in \cite{SS} to describe the next order fluctuations around $\rho^{\otimes N}$, however it does not provide estimates in the form that we need here (cf. \cite[Section 3]{RSY} for more details).

The mean field free energy functional is obtained by taking a trial state $\rho ^{\otimes N}$ in the $N$-body free energy, which yields
\begin{multline}\label{meanfieldfunct}
\MFmf [\rho] = \intR   \rho\, W_{N,m} - 2 \iint_{\mathbb R^2\times \mathbb R^2}\rho(z)\log|z-z'|\rho(z')\\ + N ^{-1} \int_{\R ^2} \rho \log \rho 
\end{multline}
with
\begin{equation*}
W_{N,m} (z) = |z| ^2 - 2 \frac{m}{N} \log |z|.
\end{equation*}
It has a unique minimizer $\rhoMFm$ among probability measures on $\mathbb R^2$.  The basic result concerning its relation to the scaled single-particle probability density
\beq \muNmone (z)= \int_{\R ^{2(N-1)}} \muNm (z,z_2,\ldots,z_N) \mathrm dz_2\ldots \mathrm dz_N\eeq 
is as follows:

\begin{teo}[\textbf{Plasma analogy for QH trial states}]\label{teo:plasma}\mbox{}
There exists a constant $C>0$ such that for large enough $N$ and any smooth function $V$ on $\mathbb R^2$
{\begin{multline}\label{densitydiff}
\left\vert \intR \left(\muNmone(z) - \rhoMFm(z)\right) V(z)\mathrm dz \right\vert \\ \leq  C\, (\log N/N)^{1/2} \Vert \nabla V \Vert_{L ^2 (\R ^2)} + C N ^{-1/2}\Vert \nabla V \Vert_{L ^{\infty} (\R ^2)}
\end{multline}}
if $m\lesssim N ^2$, and  
{\begin{multline}\label{densitydiff2}
\left\vert \intR \left(\muNmone(z) - \rhoMFm(z)\right) V(z)\mathrm dz \right\vert \\ \leq  CN^{-1/2} m ^{-1/4} \Vert V \Vert_{L ^{\infty} (\R ^2)}
\end{multline}}
if $m\gg N ^2$.
\end{teo}

The different form of the estimates \eqref{densitydiff} and \eqref{densitydiff2} are due to different methods of proofs in the two different regimes and already hint at the fact that something special happens in the regime $m\propto N ^2$, as we discuss in more details in Section \ref{sec:mean field density} below.

Essential ingredients in the proof of Theorem \ref{teo:plasma} are
\begin{itemize}
\item  2D versions of two classical electrostatic results: Onsager's lemma, and an estimate of the change in electrostatic energy when charges are smeared out (see e.g. \cite{LieSei}, Ch. 6).
\item the variational equation associated to \eqref{meanfieldfunct} and the positivity of relative entropies.
\end{itemize}
These put together give precise estimates on the $N$-body free energy functional. Using also the Csziz\`ar-Kullback-Pinsker inequality and a new refinement of Onsager's lemma one eventually deduces estimates on the one-body density. We refer to \cite{RSY} for more details.

In order to apply the theorem to estimate the energy in a potential like $V_{\omega,k}(z)=\omega |z|^2+k|z|^4$, that increases at infinity it is necessary to study also the decrease of the mean field density at infinity to compensate for the growth of $V_{\omega,k}$.

\subsection{Properties of the mean field density}\label{sec:mean field density}

Dropping the last term in \eqref{meanfieldfunct} (the entropy term) one obtains a simplified functional
\begin{multline}\label{elfunct}
\MFel [\rho] = \intR  \rho W_{N,m}  - 2 \iint \rho(z)\log|z-z'|\rho(z')
\end{multline}
that can be minimized explicitly. In fact, using the variational equation of the minimization problem and electrostatic arguments, in particular Newton's theorem, it is not difficult to prove that the minimizer, denoted by 
$\rhoMFel$, takes the constant value $1/(2\pi)$ in an annulus with inner radius $R^-=\sqrt{m/N}$ and outer radius $R^+=\sqrt{2+(m/N)}$, and is zero outside the annulus. In particular, for $m=0$ the support is a disc with radius $\sqrt{2}$. Note that the length scales here refer to scaled variables, cf. \eqref{scaledmu}. In the original variables the lengths have to be multiplied by $\sqrt N$ and the density by $N^{-1}$.

The reason for the label ``el" in \eqref{elfunct} is that the functional and the shape of the minimizer $\rhoMFel$ is entirely determined by the electrostatic terms in the free energy functional. This minimizer  can be shown to approximate 
$\rhoMFm$ in the metric that is defined by the inverse of the 2D Laplacian: Define for real valued functions (or measures) $f$ and $g$
\beq 
D(f,g)=-\iint_{\mathbb R^2\times \mathbb R^2} f(z)\log|z-z'|g(z')\,\mathrm dz\, \mathrm dz'.
\eeq
Then $D(f,f)\geq 0$ if $\int f=0$ with equality only if $f=0$. The approximation result for $\rhoMFm$ in terms of $\rhoMFel$ is
\begin{equation}
D(\rhoMFm-\rhoMFel, \rhoMFm-\rhoMFel)\leq CN^{-1}.
\end{equation}
An analogous estimate of the difference between $\muNmone$  and  $\rhoMFm$ lies behind \eqref{densitydiff}. Combining these two estimates one sees that \eqref{densitydiff} holds with $\rhoMFm$ replaced by $\rhoMFel$.
 
From the formulas for the inner and outer radii $R^\pm$ it is clear that the shape of the $\rhoMFel$ depends essentially on the ratio $m/N$. As long as $m\ll N$ the support is a disc of radius $O(1)$ with a small hole around the center. As the ratio increases the hole becomes larger and for $m\gg N$ the support is an annulus of mean radius $R \sim \sqrt{m/N}\gg 1$ and thickness $R^+-R^- \sim \sqrt {N/m}\ll 1$.
The exponential decrease of $\rhoMFm$ for large distances $r$ from the origin sets in for $r-R^+\gg \max \{R^+-R^-, N^{-1}\}$ (again we are stating the results in terms of scaled variables, cf. \eqref{scaledmu}).

The length scale $N^{-1}$ is the coefficient in the entropic term of the free energy functional and it becomes larger than the  length scale $R^+-R^-$ when $m\gtrsim N^2$. In fact, under this condition it is better to approximate $\rhoMFm$ by dropping the interaction term $D(\rho,\rho)$ from the energy functional, defining 
\begin{equation}\label{thermalfunct}
\MFth [\rho] = \intR   \rho\, W_{N,m} + N ^{-1} \int_{\R ^2} \rho \log \rho 
\end{equation}
where `$\rm th$' stands for `thermal'. The minimizer of this simplified functional, denoted $\rhoMFth$,  can be computed explicitly and is given by 
\beq  \rhoMFth(z)=(2N^{m+1})(\pi m!)^{-1}|z|^{2m}\exp(-N|z|^2).\eeq
In the radial variable $r=|z|$ this density is approximately Gaussian around $r=\sqrt{m/N}$ with width $1/\sqrt{N}$. 

This transition from an essentially electrostatic to an essentially thermal behavior of the minimizer of the mean-field functional \eqref{meanfieldfunct} and thus, via \eqref{densitydiff}, of the one-body density of Laughlin's quasi-hole when $m\gg N ^2$ does not seem to have been noticed before. Indeed, as far as we know, the usual approximations in the literature have been based on purely electrostatic arguments, that is, implicitly, on the simplified functional \eqref{elfunct}. Our analysis shows rigorously that this approximation is correct as long as $m\ll N ^2$, but breaks down in the opposite case where one should rather consider \eqref{thermalfunct}.
 
Physically, the reason for this is that to minimize \eqref{meanfieldfunct}, the density has to be squeezed as tightly as possible around the potential well of $W_{N,m}$. There are two limiting effects for the squeezing, one associated with the electrostatic term in \eqref{meanfieldfunct} and the other with the entropy term. Each term comes with its own length scale and the true density is spread over the larger of the two, which is the `electrostatic' length $\sqrt{N/m}$ for $m\ll N ^2$ and the `thermal' length for $m\gg N ^2$, indicating a transition in the regime $m\propto N ^2$. 

\medskip 

Computing the potential energy with the density $\rhoMFel$ and optimizing over $m$ leads to the same optimal value $\mopt$ as in \eqref{eq:intro m opt}. In particular, the Laughlin state, i.e., $m=0$, leading to a full disc instead of an annulus, is favored for $\omega\geq -2kN$.
Moreover, together with the estimates on the decay of $\rhoMFm$ away from the support of $\rhoMFel$, the energy upper bound \eqref{25} is improved to
\beq E_0\leq \omega N^2+\frac 43 kN^3\eeq
and \eqref{26} to
\beq E_0\leq -\frac {\omega^2N}{4k}+ \left(\frac 13k N^3+\frac 32|\omega|N\right)(1+o(1))\eeq
at least for $|\omega|/k\ll N^{7/5}\ln N$. (This limitation is due to the fact that the the current  estimates of the decay of $\rhoMFm$ are less than optimal.) Corresponding improvements hold for the angular momentum estimates: Eq. \eqref{16} is replaced by
\beq L_0\leq(4/3)^{1/2}N^2\eeq
and the $\sqrt 3$ in Eqs. \eqref{17} and \eqref{19} can be replaced by $1/\sqrt 3$.

Similar improvements of the angular momentum and energy estimates can be provided in the thermal regime using the density $\rhoMFth$, but again due to lack of precision of the decay estimates, they can only be proved with our method when $|\omega|/k\gg N ^{10/3}$.

\section{Discussion}\label{sec:conclu}

We have studied a rotating Bose gas in a quadratic plus quartic trap where the rotational frequency can exceed the frequency of the quadratic part of the trap.
Through the analysis of  trial states for energy upper bounds and simple lower bounds we have obtained criteria for the ground state to be strongly correlated in an asymptotic limit. The lower bounds, although not sharp, are of the same order of magnitude as the upper bounds.
The optimal trial state changes from a pure Laughlin state, with essentially constant density in a disc, to a modified Laughlin state with a `hole' in the density around the center when $\omega$ is negative and $|\omega|/2k$ exceeds $N$. As $|\omega|/k$ increases, the density becomes concentrated in an annulus of much smaller width than its radius. Within the annulus the density profile is still approximately flat until another transition occurs for 
$|\omega|/k$ of the order $N^2$  when the density profile becomes approximately Gaussian in the radial variable.

We believe that these results are sharp in the sense that our trial states, the Laughlin quasi-holes (Laughlin times giant vortex), are good approximations to the true ground states of \eqref{hprime} in the fully correlated regimes described in Theorem \ref{teo:correl}. We have no proof of this for the moment: Rigorous lower bounds to the energy that would match our upper estimates would require a better understanding of the properties of general fully correlated states.

Finally we discuss the connection between our approach and the experimental proposal of \cite{RRD}. There it is argued that a convenient way of creating the Laughlin state in a cold Bose gas would be via a dynamical procedure. As in our approach an anharmonic trap is used, but this only in a first step. The main idea of \cite{RRD} is to first use a ``Mexican hat'' anharmonic potential to create a rotating BEC with a large angular momentum concentrated in a giant vortex at the center of the trap. In a second step one would then stop the rotation and change adiabatically the trap from Mexican hat like to purely harmonic. With a careful tuning of parameters, one then ends up with a gas where interactions dominate the physics and thus strongly correlated states should emerge. Since the angular momentum is conserved during the adiabatic evolution, if one manages to first create a condensate with momentum $N(N-1)$, the final evolved state should be the Laughlin state, and \cite{RRD} mostly emphasizes this case. If 
one instead starts from a BEC with a larger momentum $N(N-1) + mN$ it is likely that the final evolved state will be the corresponding Laughlin times giant vortex \eqref{eq:quasi-hole}. 

Indeed, as explained in Section \ref{sec:correl states}, an interpretation of the Laughlin state is that vortices are bound to particles in order to decrease the interaction energy. In this perspective, one may interpret the adiabatic evolution proposed in \cite{RRD} as a redistribution of the vortices, all located at the center of the trap in the  initial BEC state. During the evolution, vortices are bound to particles until the interaction energy is zero. If one starts with more vortices than necessary to achieve this, i.e. with an angular momentum strictly larger than $N(N-1)$, the `excess' vortices will probably have a tendency to stay at the center of the trap to conserve rotational invariance, which leads to the form \eqref{eq:quasi-hole} as a guess for the final evolved state. A possible way to further enhance this effect and stabilize the giant vortex  could be not to completely turn off the anharmonic trap, so that the final trap also has some Mexican hat shape that would help pin the excess 
vortices at the origin.



\bigskip

\noindent{\small \textbf{Acknowledgments.}
NR thanks Xavier Blanc and Mathieu Lewin for helpful discussions during the early stages of this project. SS is supported by an EURYI award. Funding from the CNRS in the form of a PEPS-PTI project is also acknowledged.
JY thanks the Institute Mittag Leffler for hospitality during his stay in the fall of 2012.}

\end{document}